\documentclass{Interspeech}

\interspeechcameraready

\title{GigaAM: Efficient Self-Supervised Learner for Speech Recognition}

\author[]{Aleksandr}{Kutsakov}
\author[]{Alexandr}{Maximenko}
\author[]{Georgii}{Gospodinov}
\author[]{Pavel}{Bogomolov}
\author[]{Fyodor}{Minkin}

\affiliation[nocounter]{}{SaluteDevices}{Russia}
\email{\{askutsakov, ae.maximenko, georgygospodinov, bobrosoft98, minkin.fyodor\}@gmail.com}

\usepackage{comment}
\usepackage{multirow}
\usepackage{float}

\begin{document}

\maketitle

% the abstract here must exactly match the abstract entered into the paper submission system
\begin{abstract}
Self-Supervised Learning (SSL) has demonstrated strong performance in speech processing, particularly in automatic speech recognition. In this paper, we explore an SSL pretraining framework that leverages masked language modeling with targets derived from a speech recognition model. We also present chunkwise attention with dynamic chunk size sampling during pretraining to enable both full-context and streaming fine-tuning. Our experiments examine scaling with respect to model size and the amount of data. Using our method, we train the GigaAM family of models, including a state-of-the-art model for Russian speech recognition that outperforms Whisper-large-v3 by $50\%$. We have released our foundation and ASR models, along with the inference code, under the MIT license as open-source resources to the research community. Available at https://github.com/salute-developers/gigaam.
\end{abstract}

\section{Introduction}

% Self-supervised learning (SSL) has revolutionized automatic speech recognition (ASR) by leveraging vast amounts of unlabeled data, thereby reducing the need for expensive manual annotation. Numerous SSL approaches have emerged in recent years \cite{hubert, wavlm, bestrq, pbert, wav2vec2, wav2vec, data2vec}, yet most are benchmarked on narrow-domain datasets (e.g., LibriSpeech \cite{librispeech}, LibriLight \cite{librilight}) that do not fully reflect real-world conditions. In production-scale settings, it is critical not only to achieve high performance but also to ensure scalability.

Self-supervised learning (SSL) has transformed ASR by leveraging vast amounts of unlabeled data, reducing the need for manual annotation. Numerous methods have emerged \cite{hubert, wavlm, bestrq, pbert, wav2vec2, wav2vec, data2vec}, but some of them suffer from training instability due to representation collapse \cite{hubert, wav2vec2, data2vec}. On the other hand, BEST-RQ \cite{bestrq} has proven effective due to its use of a frozen audio quantization module that provides targets for masked language prediction in a single stage. However, despite its training stability, BEST-RQ suffers from semantically uninformative targets because they are derived from low-level representations. Layer-wise analyses in the literature suggest that the most semantically rich features appear in the intermediate layers, while the final layers tend to exhibit an autoencoder-like behavior \cite{layer_analysis, best_rq_autoencoder}.

%yet most are evaluated on narrow-domain datasets (e.g., LibriSpeech \cite{librispeech}, LibriLight \cite{librilight}) that do not reflect real-world conditions. For production-scale applications, models must robustly handle large data volumes without training instability or excessive hyperparameter tuning.

% Numerous methods have emerged \cite{hubert, wavlm, bestrq, pbert, wav2vec2, wav2vec, data2vec}, primarily evaluated on narrow-domain datasets (e.g., LibriSpeech \cite{librispeech}, LibriLight \cite{librilight}). However, these approaches often suffer from training instability or representation collapse and require extensive hyperparameter tuning.

Motivated by these observations, we propose an SSL pretraining framework that overcomes these limitations. Our approach leverages semantically enriched target variables derived from a supervised ASR model. Furthermore, to address potential model degeneracy on excessive silence in the audio data, we preprocess it by removing segments with insufficient speech activity. In contrast to PBERT \cite{pbert}, which extracts targets from an intermediate layer, we perform K-means clustering on the last layer of a CTC-based ASR model \cite{ctcloss} to generate targets that capture higher-level semantic information.

In addition, we explored chunk size sampling during both pretraining and fine-tuning stages and found that a single pretraining run is sufficient to support both full-context and streaming ASR setups.
% In addition, we introduce a dynamic chunk training strategy that decouples the pretraining and fine-tuning scenarios. This allows a single pretraining run to support both full-context and streaming ASR setups, thereby improving robustness and efficiency.
Using our framework, we developed and publicly released the pretrained model with its fine-tuned variants for Russian ASR. Our models achieve state-of-the-art performance on Russian speech recognition across several benchmarks \cite{golos, mcv, ruls}, outperforming competing models such as Whisper-large-v3\footnote{with forced Russian language token} by nearly $50$\%, see Table \ref{tab:main}.

In summary, our key contributions are:

\begin{itemize}
    \item \textbf{The development and public release\footnote{https://github.com/salute-developers/gigaam} of a state-of-the-art Russian ASR model}, achieving substantial performance improvements over existing approaches.

    \item \textbf{A scalable self-supervised learning (SSL) framework} that utilizes semantically enriched targets generated by a CTC-based ASR model.

    \item \textbf{A dynamic chunking strategy} enabling unified pretraining for both full-context and streaming ASR within a single run.

    \item \textbf{A thorough scalability analysis} exploring the impact of labeled/unlabeled dataset sizes and model dimensions on performance.
\end{itemize}

\begin{table}[]
    \centering
    \caption{Open-source models Word Error Rate (WER) on the Russian language datasets}
    \begin{tabular}{p{2.9cm} p{1.1cm} p{1.2cm} p{1.7cm}} %{lrrr}
    \toprule
    \textbf{Model} & \textbf{Golos Farfield} & \textbf{Russian MCV-19}  & \textbf{Russian \ \ LibriSpeech} \\
    \midrule
    XLS-R \cite{xlsr} finetuned  & 15.7 & 8.0 & 16.1 \\
    Whisper-large-v3 & 16.6 & 5.5 & 9.5 \\
    FastConformer-RNNT & 6.6 & 5.7 & 11.3 \\
    Ours (CTC) & 4.3 & 3.1 & \textbf{5.5} \\
    Ours (RNNT) & \textbf{3.9} & \textbf{2.7} & \textbf{5.5} \\
    \bottomrule
    \end{tabular}
    \label{tab:main}
\end{table}

\section{Related Work}

The field of self-supervised learning for speech encompasses a wide range of methods, from early approaches such as wav2vec \cite{wav2vec} to more recent advances. Although many excellent techniques have been proposed, in this work we focus on two representative frameworks -- HuBERT \cite{hubert} and BEST-RQ -- due to their complementary strengths. HuBERT, which forms the basis of our approach, employs a multi-stage masked language modeling pipeline to learn robust representations. At the same time, BEST-RQ is widely used in production-scale systems for its efficient, single-stage pretraining via a frozen audio quantizer \cite{usm}. In the following subsections, we detail these two approaches and discuss how our method builds upon and extends their core ideas.

\subsection{SSL Pretraining: HuBERT vs. BEST-RQ}

In the HuBERT framework, discrete target tokens are generated through a multi-stage pipeline. Given an input sequence of audio features,
% \[
% \mathcal{X} = \{ x_1, x_2, \dots, x_T \},
% \]
a subset of time indices is selected for masking.
% \[
% \mathcal{M} = \{ t_1^{mask}, t_2^{mask}, \dots, t_m^{mask} \}.
% \]
At these masked positions, the input features are replaced with a predefined mask embedding. In Stage 1, K-means clustering is applied to low-level features (e.g., MFCCs) to assign each frame a token corresponding to its nearest centroid. In Stage 2, the clustering is repeated on the intermediate representations of a pretrained (teacher) model to yield refined assignments. By contrast, BEST-RQ employs random vector quantization with frozen codebooks and projection module. This approach avoids representation collapse issues observed in jointly trained models such as wav2vec2 \cite{wav2vec2} and w2v-BERT \cite{w2v-bert}.

\subsection{Importance of Data Domain in Pretraining}

The pre-training data selection process is critical for downstream ASR performance. Prior work has shown that domain-specific models can outperform multilingual ones. For example, Babu et al. \cite{xlsr} demonstrated that monolingual models yield superior performance on domain-specific tasks, and Lu et al. \cite{google_data_selection} reported that filtering the pretraining data to match the target domain can lead to up to a $15\%$ relative reduction in Word Error Rate (WER). Motivated by these findings, our approach pretrains exclusively on Russian speech to ensure that the learned representations are optimally adapted to the target domain.

\subsection{Chunkwise Attention Mechanisms}

In practice, ASR models are often trained on relatively short audio segments (e.g., up to $30$ seconds), which can be problematic when operating in a long-form scenario. For instance, a full-context model trained on $30$-second audio fragments might struggle when presented with a $2$-minute audio file. To address this, we employ a chunkwise attention mechanism that divides the input into fixed-length segments.

Formally, given an input sequence
% \[
% \mathcal{X} = \{ x_1, x_2, \dots, x_T \},
% \]
and a predefined chunk size (e.g., $200$ms, $1$s, $2$s, $4$s, or $8$s), we partition the sequence into consecutive segments of the selected length.
% \[
% \mathcal{X}_c^{(i)} = \{ x_{(i-1)c+1}, x_{(i-1)c+2}, \dots, x_{ic} \}.
% \]
Within each chunk, attention is computed locally, enabling the model to focus on limited context rather than the entire sequence. This not only helps in generalizing to longer audio inputs but also supports streaming scenarios where only limited future context is available. We further incorporate dynamic chunk size sampling during pretraining so that the model learns to adapt to varying context lengths, allowing it to perform effectively in both full-context and streaming setups without requiring expensive re-training runs.

\subsection{Scaling Laws in SSL Pretraining}

Scaling laws have been extensively studied in natural language processing \cite{Kaplan2020, Hoffmann2022} and computer vision \cite{Dosovitskiy2021}, yet systematic investigations for self-supervised ASR remain limited. While previous works \cite{google_10b, fair_scaling, whisper} has examined the influence of model capacity, the amount of training data, and compute budget in multilingual and semi-supervised settings, a comprehensive analysis for self-supervised pretraining is still needed. In this work, we analyze scaling effects by varying the number of pretraining steps, the pretraining dataset size (spanning two orders of magnitude), the fine-tuning dataset size (spanning three orders of magnitude), and the model capacity (spanning 1.5 orders of magnitude), and we observe how these factors influence the performance of our SSL model on downstream task.

\section{Our methodology}

\subsection{Pretraining method}

% todo
% 0. intermediate layers probing
% 1. instead of hidden units we are able to take last layer
% 2. maybe move model/steps/data configurations here

\begin{figure}[t]
  \centering
  \includegraphics[width=\linewidth]{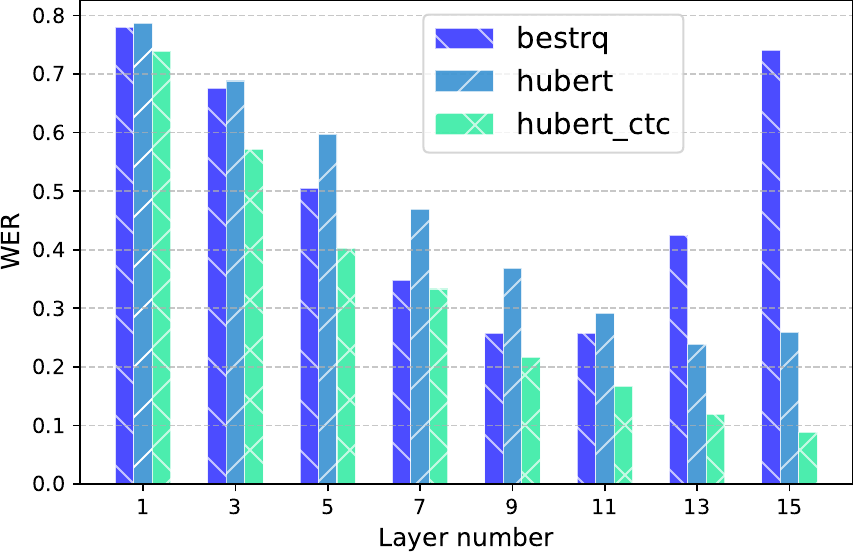}
  \caption{Intermediate layers probing of pretrained encoders with different pretraining approaches}
  \label{fig:probing}
\end{figure}

Our pretraining method HuBERT-CTC builds on HuBERT, where KMeans clustering is applied to the hidden states of the encoder. However, unlike the standard HuBERT approach, which uses a pretrained encoder, we employ an encoder fine-tuned on ASR tasks. We compare the properties of these approaches by probing the intermediate layers of the pretrained encoders on speech recognition. As shown in Fig. \ref{fig:probing}, the BEST-RQ/HuBERT pretrained encoders achieve the best performance at the intermediate layers, with quality degrading towards the last layers. In contrast, our method demonstrates a monotonic improvement in ASR performance up to the latest layer. This results in more ASR-specific representations being learned, which is particularly crucial in low-resource settings.

\subsection{Data preprocessing}

\subsubsection{Voice Activity Detection filtering}

To improve the quality of our pretraining data, we employ Voice Activity Detection (VAD) filtering to identify and remove segments with excessive silence, similar to \cite{assembly_conformer, usm}. Each audio sample from the pretraining dataset was implicitly split into one-minute chunks to avoid discarding large samples during filtering. The segments in which the proportion of silence exceeded $60\%$ were discarded. This process retained $80\%$ of the original data while improving the model performance by an average of $5-10\%$.

\subsubsection{More advanced data filtering methods}

In addition to VAD filtering, we explore more sophisticated data filtering techniques, including methods from SemDeDup \cite{semdedup} and data pruning \cite{ssl_pruning}. These algorithms were originally designed for language models and aim to reduce redundancy and improve the diversity of training data. We applied them both separately and on top of VAD-filtered data, however, the observed changes in model performance were negligible. Given the significant computational overhead of these methods and their limited impact, we decided to rely solely on VAD filtering for our data preprocessing pipeline.

\subsection{Streaming and long-form inference}

Chunkwise attention enables the model to limit its context size and process audio of varying lengths, different from those seen during training. By using a sufficiently large chunk size of several seconds, the receptive field can extend to approximately a minute. Streaming applications are achieved by reducing the chunk size and employing chunkwise causal convolutions \cite{chunk_causal} in the conformer layers, ensuring that the required future context is limited to the chunk size and allowing the reuse of pretrained non-streaming version.

\section{Experiments}

% todo
% 0. make less `with a` (?)
% 1. (!!) teacher model pretraining method and hours

Our experimental setup is as follows: for self-supervised pretraining with HuBERT-CTC, we use $100$k hours of Russian audio and train for $400$k steps with a virtual batch size of $9$ hours. For supervised fine-tuning, we utilize datasets including Golos \cite{golos}, Mozilla Common Voice (ru) \cite{mcv}, Russian LibriSpeech \cite{ruls}, and SOVA \cite{sova}, totaling $2$k hours of audio, and train for $100$k steps with a batch size of $2048$ samples. Across all experiments, we employ the Conformer \cite{conformer} architecture with a standard model size of $240$M parameters. The $240$M teacher model for HuBERT is trained on $50$k hours of pretraining audio using default hyperparameters and the wav2vec2 approach. HuBERT-CTC utilizes a fine-tuned version of this model with the same supervised data as the other models in the paper. To compute the final WER in the following experiments, we average the metrics over the last ten epochs on the test sets (Golos Farfield/Crowd, MCV, LibriSpeech). For few-shot supervised data, where overfitting is likely, we employ early stopping.

\subsection{Other pretraining methods}

% todo
% 1. add final WER calculation for this case
% 4. (!!) using scenario - distillation large ASR model with a small FT
%   - hubert requires intermediate layers (?) outputs (probing, etc)
%   - ours works with the last layer outputs and achieves faster convergence

To compare our approach with other pretraining methods, including the setup without any pretraining at all, we finetune the model with different amounts of supervised data ($100\%$, $10\%$, $1\%$, $0.1\%$). Our results are presented in Table \ref{tab:pre_methods}. Even when using the full dataset, we observe a significant performance gap between our method and others, which further widens as the amount of supervised data decreases. This demonstrates the ability of HuBERT-CTC to distill knowledge from the ASR model even on unlabeled data, highlighting its applicability in low-resource scenarios.

\begin{table}[th]
  \caption{The final WER for different pretraining methods and fine-tuning set sizes}
  \label{tab:pre_methods}
  \centering
  % \sisetup{table-format=2.2} % Set the format for numbers in the table
  \begin{tabular}{ lrrrr}
    \toprule
    % \multirow{2}{*}{\textbf{Pre-train}}
     & \multicolumn{4}{c}{\textbf{Fine-tuning set size}} \\
    \cmidrule(lr){2-5}
    \textbf{Pretraining} & \textbf{100\%} &
    \textbf{10\%} & \textbf{1\%} & \textbf{0.1\%} \\
    \midrule
    w/o pretraining & 5.71 & 21.12 & 76.78 & {--} \\
    BEST-RQ & 5.18 & 6.47 & 9.90 & 16.80 \\
    Wav2vec2 & 4.62 & 6.6 & 10.66 & 19.38 \\
    HuBERT & 3.66 & 5.19 & 8.48 & 16.70 \\
    HuBERT-CTC & 3.35 & 4.44 & 5.77 & 7.41 \\
    \bottomrule
  \end{tabular}
\end{table}

% \begin{table*}[th]
%   \caption{The final WER for different pretraining methods, fine-tuning set sizes, and ASR decoders}
%   \label{tab:pre_methods}
%   \centering
%   \sisetup{table-format=2.2} % Set the format for numbers in the table
%   \begin{tabular}{ l *{8}{S} }
%     \toprule
%     & \multicolumn{2}{c}{\textbf{100\% FT (2000 h)}} &
%     \multicolumn{2}{c}{\textbf{10\% FT (200 h)}} &
%     \multicolumn{2}{c}{\textbf{1\% FT (20 h)}} &
%     \multicolumn{2}{c}{\textbf{0.1\% FT (2 h)}} \\
%     \cmidrule(lr){2-3} \cmidrule(lr){4-5} \cmidrule(lr){6-7} \cmidrule(lr){8-9}
%     \textbf{pretraining method} & {\textbf{CTC}} & {\textbf{RNNT}} & {\textbf{CTC}} & {\textbf{RNNT}} & {\textbf{CTC}} & {\textbf{RNNT}} & {\textbf{CTC}} & {\textbf{RNNT}} \\
%     \midrule
%     w/o pretraining                  &  5.71 &  5.65 & 21.12 & 20.85 & 76.78 & 71.89 & {--} & {--} \\
%     Best-RQ            &  5.18 &  4.12 &  6.47 &  6.07 &  9.90 &  9.61 & 16.80 & 16.17 \\
%     HuBERT             &  3.66 &  3.51 &  5.19 &  4.79 &  8.48 &  8.47 & 16.70 & 15.42 \\
%     HuBERT-CTC (ours)  &  3.35 &  3.04 &  4.44 &  4.35 &  5.77 &  5.73 &  7.41 &  7.55 \\
%     % HuBERT-RNNT (ours) &  3.25 &  2.98 &  4.47 &  4.08 &  6.52 &  6.25 & 11.63 & 10.07 \\
%     \bottomrule
%   \end{tabular}
% \end{table*}

\subsection{Data, model and steps scaling}

We conduct a comprehensive investigation into the relationship between model performance and the amount of (un)labeled data, model capacity, and computational budget. To provide clarity, we outline all configurations used in our experiments:
\begin{itemize}
    \item Pretraining dataset size: $1$k, $6$k, $12$k, $24$k, $50$k, $100$k hours.
    \item Fine-tuning dataset size: $2000, 200, 20, 2$ hours.
    \item Model size: $30$M, $60$M, $100$M, $240$M, $500$M.
    \item Number of pretraining steps: $25k, 100k, 400k$ steps.
\end{itemize}

\subsubsection{Pretraining data}

% todo
% 1. maybe add scaling 1k pretraining steps to the graph

\begin{figure}[t]
  \centering
  \includegraphics[width=\linewidth]{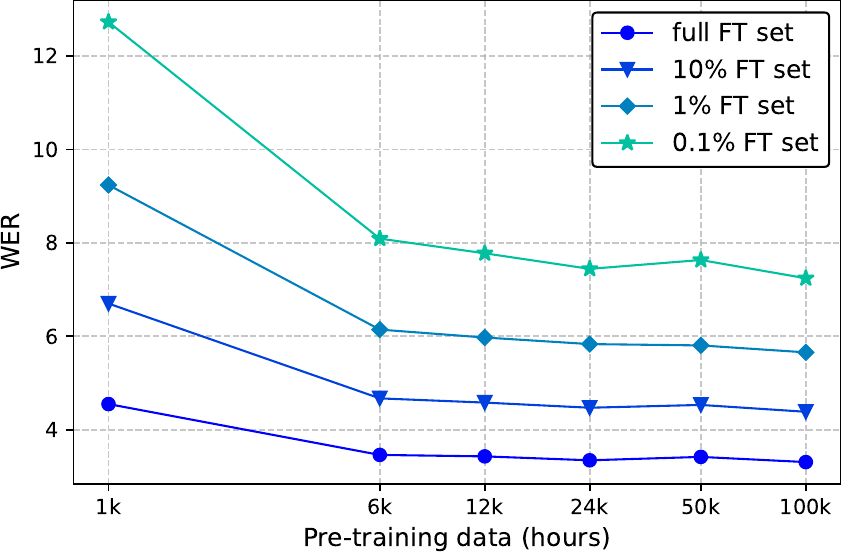}
  \caption{The dependency of WER on the pretraining dataset size with fine-tuning set scaling}
  \label{fig:data_scaling}
\end{figure}

The reduction of supervised data quantity significantly affects speech recognition quality, but the dependence on the amount of unsupervised data is less frequently explored. The results of our investigation, spanning from $1$k to $100$k hours of unsupervised audio, are presented in Figure \ref{fig:data_scaling}. A key property of the HuBERT-CTC approach is the stabilization of model performance after reaching a certain threshold of pretraining data. Specifically, for values above $6$k hours, the final quality remains nearly unchanged, as the student model effectively transfers knowledge from the teacher even with a relatively small dataset. This property will be particularly advantageous for distilling large multilingual models into smaller, domain-specific ones, and we leave further exploration of this direction as future work. Notably, a significant degradation in quality occurs only when the unsupervised data is reduced to just $1$k hours, which is half the size of the supervised dataset.

\subsubsection{Model size}

\begin{figure}[t]
  \centering
  \includegraphics[width=\linewidth]{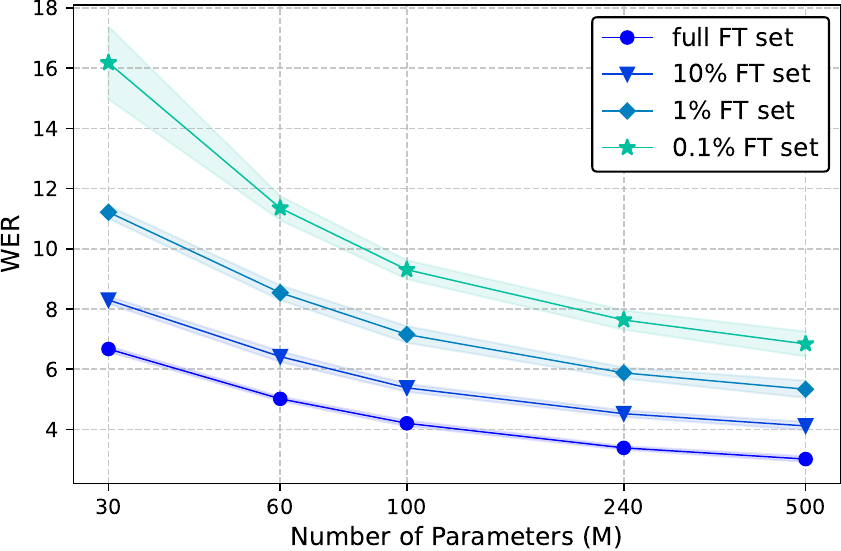}
  \caption{Quality dependency on model size with fine-tuning set scaling. For each configuration, we average the metrics across pre-training data quantities ranging from $6k$ to $100k$ hours and include the standard deviation on the graph}
  \label{fig:dim_scaling}
\end{figure}

We investigate the relationship between model performance and size, exploring models ranging from $30$M to $500$M parameters. The amount of supervised data is varied similarly to the previous experiment. The results are presented in Figure \ref{fig:dim_scaling}. Since varying the amount of pretraining data from $6$k to $100$k hours yields only minor changes in performance, we compute the mean across these five configurations and include the standard deviation on the graph. Remarkably, a $100$M model outperforms the $240$M teacher model ($4.62 \to 4.21$ WER), even when pretrained on a smaller amount of data.

\subsubsection{Pretraining as regularization}

% todo
% 1. we find it useful in no-pretrain cases

We also explored fine-tuning regularization techniques aimed at accelerating convergence and improving model generalization. Among these, the CR-CTC \cite{crctc} method showed particular promise, significantly enhancing the final performance of smaller models in the absence of pretraining. Table \ref{tab:crctc} presents the results in terms of relative WER improvement (in percentage) after applying CR-CTC for different model sizes and pretraining durations. Notably, the application of CR-CTC does not benefit larger models, while for smaller models, pretraining effectively replaces the need for regularization, resulting in only marginal changes when CR-CTC is applied.

\begin{table}[ht]
    \centering
    \caption{CR-CTC impact comparison with partial pretraining, given as a percentage improvement w/o $\to$ w/ CR-CTC. The mean and standard deviation are calculated across pretraining data quantities ranging from $6k$ to $100k$ hours. The column labeled ``{--}'' corresponds to training from scratch without pretraining}
    \begin{tabular}{lrrrr}
    \toprule
    \multirow{2}{*}{\textbf{Model}} & \multicolumn{4}{c}{\textbf{Pretraining steps}} \\
    \cmidrule(lr){2-5}
     & \textbf{{--}} & \textbf{25k} & \textbf{100k}  & \textbf{400k} \\
    \midrule
    $30$M  &  $17.0$ &  $15.7 \pm 0.7$ &   $3.0 \pm 0.5$ &   $1.5 \pm 0.4$ \\
    $60$M  &  $14.2$ &  $13.4 \pm 3.8$ &   $2.0 \pm 1.0$ &  $-0.7 \pm 1.3$ \\
    $100$M &  $13.5$ &   $9.6 \pm 1.8$ &   $3.3 \pm 1.8$ &  $-0.6 \pm 0.5$ \\
    $240$M &   $4.4$ &  $-3.3 \pm 1.4$ &  $-0.4 \pm 1.8$ &  $-3.1 \pm 0.4$ \\
    $500$M &  $-6.7$ &  $-3.3 \pm 1.1$ &   $0.1 \pm 0.8$ &   $0.6 \pm 0.9$ \\
    \bottomrule
    \end{tabular}
    \label{tab:crctc}
\end{table}

\subsection{Streaming and long-form inference}

In our experiments we utilize chunkwise attention with chunk sizes of $200$ms, $1$s, $2$s, $4$s, and $8$s, as well as dynamic chunking, where chunk sizes were uniformly sampled from these values. During pretraining, a chunk size of $200$ms is excluded, as the training process tends to diverge. Table \ref{tab:chunk} presents the results of various pretraining chunking configurations fine-tuned with specific chunk sizes, including both chunkwise causal and standard convolutions. For the $200$ms chunk, only streaming convolutions are used, as this size is impractical for long-form. The dynamic chunking proves to be the most versatile approach for pretraining, consistently delivering some of the best results even when fine-tuning (and during inference) with full-context attention. We also experimented with dynamic chunk size sampling during fine-tuning, but this led to a degradation in quality.

\subsubsection{Limited kernel size}

To limit the receptive field increase to one chunk per layer, we employ a small kernel size of $5$ ($200$ms) for convolutions in our model. We conducted an ablation study with a larger kernel size of $31$ ($>1$s) and observed a slight degradation in performance ($3.35 \to 3.49$ WER) compared to the smaller kernel size. This result allows us to use a small kernel size without losing quality or unnecessarily expanding the context size.

\begin{table}[ht]
\centering
\caption{Chunking results: the ``pretraining'' column specifies the pretraining chunking method used. For convenience, results are grouped by the fine-tuning chunk size (full-context option is given by ``{--}''). ``Long-form'' refers to standard convolutions, ``streaming'' corresponds to chunkwise causal convolutions, where the future context is restricted to one chunk}
\begin{tabular}{lcccc}
\toprule
\multirow{2}{*}{\textbf{Pretraining}} & \multirow{2}{*}{\textbf{FT chunk}} & \multicolumn{2}{c}{\textbf{WER}} \\
\cmidrule(lr){3-4}
 & & \textbf{Long-form} & \textbf{Streaming}  \\
\midrule
Full-context & {--} & 3.35 & {--} \\
\midrule
Dynamic chunk & {--} & 3.31 & {--} \\
\midrule

Full-context & \multirow{3}{*}{$8$s} & $3.47$ & $3.52$ \\
Chunk=$8$s & & $3.39$ & $3.39$ \\
Dynamic chunk & & $3.39$ & $3.44$ \\
\midrule
Full-context & \multirow{4}{*}{$4$s} & $3.92$ & $4.02$ \\
Chunk=$8$s & & $3.56$ & $3.83$ \\
Chunk=$4$s & & $3.58$ & $3.85$ \\
Dynamic chunk & & $3.51$ & $3.77$ \\
\midrule
Full-context & \multirow{5}{*}{$2$s} & $4.42$ & $5.03$ \\
Chunk=$8$s & & $3.91$ & $4.66$ \\
Chunk=$4$s & & $3.83$ & $4.7$ \\
Chunk=$2$s & & $3.62$ & $4.49$ \\
Dynamic chunk & & $3.63$ & $4.56$ \\
\midrule
Full-context & \multirow{6}{*}{$1$s} & $5.09$ & $6.37$ \\
Chunk=$8$s & & $4.32$ & $5.88$ \\
Chunk=$4$s & & $4.2$ & $5.81$ \\
Chunk=$2$s & & $3.94$ & $5.56$ \\
Chunk=$1$s & & $3.94$ & $5.68$ \\
Dynamic chunk & & $3.97$ & $5.49$ \\
\midrule
Full-context & \multirow{6}{*}{$200$ms} & {--} & $12.3$ \\
Chunk=$8$s & & {--} & $11.71$ \\
Chunk=$4$s & & {--} & $10.98$ \\
Chunk=$2$s & & {--} & $10.65$ \\
Chunk=$1$s & & {--} & $10.35$ \\
Dynamic chunk & & {--} & $10.37$ \\
\bottomrule
\end{tabular}
\label{tab:chunk}
\end{table}

\section{Conclusion}

We presented an SSL pretraining framework for ASR which leverages semantically enriched targets from a CTC-based ASR model. By integrating dynamic chunk size training, our method supports both full-context and streaming fine-tuning ability within a single pretraining run. Experiments on Russian speech recognition demonstrate that our HuBERT-CTC method outperforms BEST-RQ, HuBERT, and wav2vec2 -- with our proposed models achieving state-of-the-art performance and surpassing Whisper-large-v3 by nearly $50\%$. It is also worth noting that all other presented models utilize autoregressive decoding, and many of them are significantly larger than ours. Despite this, they are outperformed by even the non-autoregressive CTC version of our model. Our scaling analysis confirms the robustness of the approach.

% todo
% using scenario - distillation large ASR model with small pretraining and fine-tuning sets (small data pretrain works fine)

% \section{Acknowledgements}
% Acknowledgement should only be included in the camera-ready version, not in the version submitted for review. The 5th page is reserved exclusively for acknowledgements and  references. No other content must appear on the 5th page. Appendices, if any, must be within the first 4 pages. The acknowledgments and references may start on an earlier page, if there is space.

% \ifinterspeechfinal
%      The Interspeech 2025 organisers
% \else
%      The authors
% \fi
% would like to thank ISCA and the organising committees of past Interspeech conferences for their help and for kindly providing the previous version of this template.

\bibliographystyle{IEEEtran}
\bibliography{mybib}

\end{document}